\documentclass[a4paper,twocolumn,fleqn]{article}

\usepackage{amsmath}             
\usepackage{txfonts}               
\usepackage{helvet}
\usepackage{bm}                  
\usepackage{cite}                  
\usepackage[dvipdfm]{graphicx}     
\usepackage{jpfr}                  

\usepackage{yfonts}
\newcommand{\gt}[1]{\mbox{\textfrak{#1}}}

\usepackage{color}
\newcommand{\change}[1]{\textcolor{red}{#1}}
\renewcommand{\change}[1]{#1}

\newcommand{\exb}{$\Vec{E} \times \Vec{B}$ }
\newcommand{\fig}[1]{Fig.\ \ref{#1}}
\newcommand{\figs}[2]{Figs.\ \ref{#1}--\ref{#2}}
\newcommand{\kpar}{k_{\parallel}}
\newcommand{\kperp}{k_{\perp}}
\newcommand{\kperpc}{k_{\perp \textrm{c}}}

\newcommand{\gyroavg}[2]{\langle #1 \rangle_{\Vec{#2}}}

\newcommand{\phik}{\varphi_{\Vec{k}}}
\newcommand{\sect}[1]{Sec.\ \ref{#1}}
\newcommand{\vexb}{v_{\Vec{E}\times\Vec{B}}}
\newcommand{\vperp}{v_{\perp}}
\newcommand{\vpar}{v_{\parallel}}
\newcommand{\vth}{v_\textrm{th}}

\renewcommand{\Vec}[1]{\bm{#1}}

\newcommand{\figdir}{figs}
\newcommand{\figsize}{6.5cm}
\newcommand{\arxiv}[1]{\textit{E-print} arXiv:#1}
\newcommand{\prl}{Phys.\ Rev.\ Lett.\ }

\begin{document}

\title{
  Gyrokinetic simulation of entropy cascade\\
  in two-dimensional electrostatic turbulence
}


\author{
  T.~TATSUNO\sup{1}, M.~BARNES\sup{1,2,3}, S.~C.~COWLEY\sup{3,4},
  W.~DORLAND\sup{1}, G.~G.~HOWES\sup{5}, R.~NUMATA\sup{1},
  G.~G.~PLUNK\sup{1,6,7}, and A.~A.~SCHEKOCHIHIN\sup{2,4}
}

\affiliation{
  \sup{1}IREAP, University of Maryland, College Park, Maryland 20742, USA\\
  \sup{2}Rudolf Peierls Centre for Theoretical Physics, University of Oxford, Oxford OX1 3NP, UK\\
  \sup{3}Euratom/UKAEA Fusion Association, Culham Science Centre, Abingdon OX14 3DB, UK\\
  \sup{4}Plasma Physics, Blackett Laboratory, Imperial College, London SW7 2AZ, UK\\
  \sup{5}Department of Physics and Astronomy, University of Iowa, Iowa City, Iowa 52242, USA\\
  \sup{6}Department of Physics and Astronomy, University of California, Los Angeles, California 90095, USA\\
  \sup{7}Wolfgang Pauli Institute, University of Vienna, Vienna A-1090, Austria
}

\date{\today}

\email{tatsuno@umd.edu}

\begin{abstract}
  Two-dimensional electrostatic turbulence in magnetized weakly-collisional plasmas exhibits a cascade of entropy in phase space [\prl \textbf{103}, 015003 (2009)].
  At scales smaller than the gyroradius, this cascade is characterized by the dimensionless ratio $D$ of the collision time to the eddy turnover time measured at the scale of the thermal Larmor radius.
  When $D \gg 1$, a broad spectrum of fluctuations at sub-Larmor scales is found in both position and velocity space.
  The distribution function develops structure as a function of $\vperp$, the velocity coordinate perpendicular to the local magnetic field.
  The cascade shows a local-scale nonlinear interaction in both position and velocity spaces, and Kolmogorov's scaling theory can be extended into phase space.
\end{abstract}


\maketitle  


\section{Introduction}

Fluid turbulence can be described mostly by Navier-Stokes equations,
which consist of partial differential equations in the position space
--- thus in 2D or 3D space.  Such a description is possible because we
may regard the fluid to be sufficiently collisional and to be locally
in thermodynamic equilibrium.  From the
outer driving scales to the
small dissipative scales, the energy (or the enstrophy in
the 2D case) cascades in the wave-number space \cite{Kolmogorov}.
The {\em inertial range} dynamics are basically governed by
the nonlinear term.  The fluxes are finally dissipated by the
viscosity described by a diffusion operator, in the {\em dissipation range}
of wave-number space.  There is a dimensionless number called
the Reynolds number, which characterizes the scale separation of the
turbulent dynamics.  The ratio of the dissipation scale to the
outer scale is in general determined by a fractional power of the Reynolds
number.

Kinetic turbulence is less well understood.
We have some knowledge especially from the extensive study of fusion plasmas
\cite{WatanabeSugama-PoP04,Idomura-PoP06,Candy-PPCF07} and recently of space
plasmas \cite{Bale-PRL05,Howes-PRL08,GarySaitoLi,Sahraoui-PRL09,Alexandrova-PRL09},
but not to the same extent as fluid
turbulence is known, particularly in the sense described above.  Since most
plasmas are in a collisionless or weakly-collisional
environment, they are not in local thermodynamical
equilibrium.  In this case we have to invoke kinetic description in
the phase space, which enables us to use Vlasov or
Boltzmann theory, or their magnetized low-frequency limit, the gyrokinetics.
Dissipation in the kinetic system occurs because of collisions.  Without
collisions, the entropy is conserved and the system is in principle
reversible in a thermodynamic sense \cite{Boltzmann}.  Wave resonances
or Landau damping do not by themselves constitute dissipation since
no entropy is generated unless we include the effect of collisions.
Irreversibility becomes the key issue here.

In this context, the following question arises.  Collisions are
usually described
by a diffusion operator in the \textit{velocity space} \cite{Helander}. How
do collisions in a weakly collisional system determine the wave-number
as well as velocity-space cutoff of the inertial range?
In this paper, following Ref.\ \cite{Tatsuno-PRL09}, we show detailed numerical evidence of the coupling between position- and velocity-space structures, and the consequent achievement of the collisional dissipation described by the velocity-space diffusion operator.
We introduce a new dimensionless number $D$, which, analogously
to the Reynolds number in Navier-Stokes equations,
characterizes the scale separation between the outer scale and the
dissipation scale in the kinetic theory.  We emphasize that we are
concerned only with scales smaller than the Larmor radius.

In order to focus on the nonlinear interaction, we omit Landau damping by ignoring variation along the field line.
The system retains two collisionless invariants as the 2D Navier-Stokes equations do.
These two invariants cannot share the same local-interaction space in a Kolmogorov-like phenomenology, and thus lead to a dual cascade (direct and inverse cascades) \cite{Idomura-PoP06,Fjortoft-Tellus53,Kraichnan-PoF67,HasegawaMima-PoF78,Horton-PoP00}.
In this paper we focus on the direct cascade only and the discussion of the inverse cascade is left for future publication.

This paper is organized as follows.
As a reduced kinetic model for plasma turbulence, we first introduce the gyrokinetic (GK) equation briefly and describe its basic nature in \sect{sec:equations}.
Scaling relations of the turbulent cascade are also briefly summarized.
Section \ref{sec:direct} shows the evidence of kinetic turbulent cascade by means of the numerical simulation of the decaying turbulence.
In addition to the reproduction of the theoretical prediction, we show how much resolution is needed in both position and velocity space to achieve the proper dissipation and show the trend that the amount of dissipation is asymptotically independent of the collision frequency.
We also present the characteristics of the nonlinear triad interaction in wave-number space and show that the assumption of local-scale interaction is supported.
We finally summarize our results in \sect{sec:summary}.

\section{Gyrokinetics}
\label{sec:equations}

\subsection{Equations}

We first introduce the GK model briefly \cite{Catto-PP78,AntonsenLane-PoF80,FriemanChen-PoF82,Howes-ApJ06}.
Since we are going to look at the turbulence in the magnetized plasmas, the dynamics of interest is much slower than particles' gyromotion.
We may average over the gyromotion and the gyroangle can be ignored due to this gyroaveraging.
A GK system has 3 spatial coordinates ($x$, $y$, $z$), and 2 velocity coordinates ($\vperp$, $\vpar$), where $\parallel$ and $\perp$ stand for parallel and perpendicular directions to the background magnetic field.
It is convenient to distinguish between the particle coordinate $\Vec{r}$ and the gyrocenter coordinate $\Vec{R}$ \cite{Catto-PP78}.
These coordinates are connected by 
\begin{equation}
  \Vec{R} = \Vec{r} + \frac{\Vec{v} \times \Vec{\hat{z}}}{\Omega},
  \label{Catto transform}
\end{equation}
where $\Vec{\hat{z}}$ is the unit vector along the background magnetic field and $\Omega$ is the gyrofrequency.

We further reduce the GK equation into 2D in position space, or 4D in phase space, by ignoring variation along the mean field ($\kpar=0$).
This is useful because we not only reduce the dimension but also remove the Landau damping from the system.
The resultant equation is
\begin{equation}
  \label{eq:gk}
  \frac{\partial h}{\partial t} +  \frac{c}{B_0} \{ \gyroavg{\varphi}{R},
  h \} = \frac{q F_0}{T_0} \frac{\partial \gyroavg{\varphi}{R}}{\partial t}
  + \gyroavg{C[h]}{R},
\end{equation}
where $h$ is the non-Boltzmann part of the perturbed ion distribution function, $\varphi$ is the electrostatic potential, $B_0$ is the background magnetic field aligned with the $z$-axis, $\gyroavg{\cdot}{R}$ is the gyroaverage holding the guiding center position $\Vec{R}$ constant, $\{f,g\}= (\Vec{\hat{z}} \times \nabla f) \cdot \nabla g$ is the standard Poisson bracket, $q$ is the charge, and $T_0$ is the temperature of the background Maxwellian $F_0$.
The collision operator $C[h]$ we use in our simulations describes pitch-angle scattering and energy diffusion with proper conservation properties (see \cite{Abel-PoP08}).

To calculate the self-consistent potential at these small scales, we use the quasineutrality condition
\begin{equation}
  \label{eq:qn}
  Q \varphi = q \int \gyroavg{h}{r} \, d\Vec{v}
\end{equation}
where $\gyroavg{\cdot}{r}$ denotes the gyroaverage at fixed particle position $\Vec{r}$, 
$Q = \sum_s q_s^2 n_{0s} / T_{0s}$ for Boltzmann-response (3D) electrons or $Q = q_i^2 n_{0i} / T_{0i}$ for no-response (2D) electrons, $n_0$ is the density corresponding to $F_0$, and $s$ and $i$ are the species indices.
In this paper we use no response electrons since electrons cannot contribute to the potential if $\kpar = 0$ exactly \cite{DorlandHammett-PoFB93}.
We have also made a separate two-species simulation with real mass ratio and obtained almost identical results in the regime $\kperp \rho_e \lesssim 1$ with negligible density fluctuations, where $\rho_e$ is the electron Larmor radius.

There are two collisionless invariants,
\begin{align}
  \label{eq:wes}
  \gt{W} &= \iint \frac{T_0 h^2}{2 F_0} \, d\Vec{R} \, d\Vec{v}
  - \frac{Q}{2} \int \varphi^2 d\Vec{r}, \\
  \label{eq:w4d}
  \gt{E} &= \frac{Q}{2} \sum_{\Vec{k}} (1 - \Gamma_0) |\varphi_{\Vec{k}}|^2,
\end{align}
where the subscript $\Vec{k}$ is for the Fourier coefficients, $\Gamma_0 = I_0(b) e^{-b}$,  $I_0$ is the modified Bessel function, $b = \kperp^2 \rho^2 / 2$, and $\rho$ is the ion thermal Larmor radius. The invariant $\gt{W}$ is proportional to the negative of the perturbed part of the entropy of the system, $-\int f \ln f \, d\Vec{r} \, d\Vec{v}$ \cite{Krommes,WatanabeSugama-PoP04}, where $f$ is the full distribution function.
Here we will refer to $\gt{W}$ as ``entropy'' to emphasize this connection, although in thermodynamic terms, it is better interpreted as a generalized free energy \cite{Hallatschek,Alex}.
The second invariant $\gt{E}$ implies the existence of a dual cascade in the 2D system, which we do not discuss in this paper.
Note that these quantities are conserved only in the collisionless limit.
Any decrease of $\gt{W}$ due to collisions corresponds to the creation of entropy and heating, hence the irreversibility.

\subsection{Turbulent scaling}
\label{sec:theory}

A scaling theory of the entropy cascade in the sub-Larmor scale range ($\ell \ll \rho$) can be developed in a way reminiscent of the Kolmogorov-style turbulence theories.
From the conservation of $\gt{W}$ in the collisionless limit, we may assume that the entropy is going to cascade scale by scale without dissipation in the inertial range.
Note that the quantity which cascades forward (to smaller scales) is $\gt{W}$, not $\gt{E}$.
This has been demonstrated by applying Fj{\o}rtoft's argument on the dual cascade \cite{Fjortoft-Tellus53,Alex}.

The nonlinear term [the Poisson bracket in \eqref{eq:gk}] that causes
the entropy transfer introduces velocity-space structure simultaneously.
For small-scale electric fields, particles with different gyroradii execute different \exb motions because they see different effective potentials; this leads to nonlinear phase mixing and other novel phenomena \cite{DorlandHammett-PoFB93,Gustafson-PoP08}.  
It can be argued that the nonlinear phase mixing
produces the following relationship between position-space
and velocity-space scales:
\begin{equation}
  \frac{\delta \vperp}{\vth} \sim \frac{\ell}{\rho}.
  \label{correlation}
\end{equation}

Dimensional arguments then lead to the following scalings of the spectra (see Refs.\ \cite{Tatsuno-PRL09,Alex,Plunk-JFM} for details),
\begin{equation}
  E_h(\kperp) \propto \kperp^{-4/3}, \quad
  E_{\varphi}(\kperp) \propto \kperp^{-10/3},
  \label{eq:spectra}
\end{equation}
where 
\begin{align}
  E_h(\kperp) &= \sum_{|\Vec{k}_\perp| = \kperp} \int \frac{T_0}{2F_0}
    |h_{\Vec{k}}|^2 \, d\Vec{v}, \label{def:h-spectra}\\
  E_{\varphi}(\kperp) &= \sum_{|\Vec{k}_\perp| = \kperp}
    \frac{q^2 n_0}{2T_0} |\varphi_{\Vec{k}}|^2. \label{def:phi-spectra}
\end{align}
Note that the total entropy \eqref{eq:wes} can be expressed as $\gt{W} = \int [E_h(\kperp) - E_{\varphi}(\kperp)] \, d\kperp$.

At small position-space scales, collisions tend to dominate as
velocity-space structure becomes finer due to \eqref{correlation}.
It can be shown \cite{Tatsuno-PRL09,Alex,Plunk-JFM} that the cutoffs are
\begin{equation}
  \frac{\delta v_{\perp{\rm c}}}{\vth} \sim \frac{1}{\kperpc\rho} \sim D^{-3/5},
  \label{cutoff scale}
\end{equation}
where 
\begin{equation}
  D = \frac{1}{\nu \tau_{\rho}},
  \label{dimensionless number}
\end{equation}
the quantity $\nu$ is the ion collision frequency and $\tau_{\rho}$ is a turnover time at the scale $\rho$.

We introduced a dimensionless number $D$ which represents the ratio between the outer scale, in this case the thermal Larmor radius, and the dissipation scale valid for the kinetic theory.
This is simply the ratio of the collision time and the eddy turnover time at the ion gyroscale.
It describes how many times eddies at the Larmor radius scale turn
over before the cumulative effect of small-angle collisions smears out
the resulting structures in velocity space at this scale.
Analogous to the Reynolds number, $D$ becomes large when
dissipation is weak, i.e., when the collision frequency is small.
We emphasize here that this number is amplitude dependent since turnover time is determined by the \exb velocity where $\Vec{E}$ is a fluctuation.

Table \ref{table of D} shows $D$ values in more realistic and geometrically complicated physical systems.
\begin{table}[bt]
  \caption{
    Estimated dimensionless number $D$ in various physical systems.
    Ion scale turbulence at LAPD \cite{CarterMaggs-PoP09},
    DIII-D \cite{Jakubowski-PRL02} and TFTR \cite{Durst-PRL93},
    ETG turbulence at NSTX \cite{Mazzucato-PRL08},
    and solar wind observation \cite{Bale-PRL05}.
    Eddy turnover time $\tau$ for the ETG case is evaluated from simulation data \cite{JenkoDorland-PRL02}.
  }
  \begin{center} \begin{tabular}{cccc}
    \hline \hline
    & $\nu$ [$\mbox{sec}^{-1}$] & $\tau$ [sec] & $D$\\\hline
    LAPD & $2.1 \cdot 10^6$ & $\gtrsim 5.0 \cdot 10^{-6}$ & $\lesssim 0.1$\\
    DIII-D & $3300$ & $7.5 \cdot 10^{-6}$ & $40$\\
    TFTR & $120$ & $1.5 \cdot 10^{-5}$ & $500$\\
    ETG @ NSTX & $10,800$ & $3.5 \cdot 10^{-7}$ & $270$\\
    Solar wind & $2.3 \cdot 10^{-7}$ & $190$ & $2.3 \cdot 10^4$\\
    \hline \hline
  \end{tabular} \end{center}
  \label{table of D}
\end{table}
The first three are the ion-scale turbulence in the basic plasma device LAPD \cite{CarterMaggs-PoP09} and 
in the fusion devices DIII-D \cite{Jakubowski-PRL02} and TFTR \cite{Durst-PRL93}.
The fourth one is the electron-temperature-gradient (ETG) driven turbulence in NSTX, where $\nu$ is estimated from experimental conditions \cite{Mazzucato-PRL08} while $\tau$ is from the streamer-dominated results of simulation \cite{JenkoDorland-PRL02} at $\kperp \rho_e = 0.2$.
The last one is for the solar wind turbulence for which the data from Cluster spacecraft at 19 earth radii is used \cite{Bale-PRL05}.
In the solar wind, the collision frequency is extremely small because of the small density, leading to extremely large $D$.

\section{Numerical simulation}
\label{sec:direct}

\subsection{Code: AstroGK}
AstroGK is an Eulerian initial value solver for the GK equations in 5D phase space in general, although we only use 4 dimensions in this study.
Namely, we discretize the velocity space into grids as standard fluid codes do in the position space.
It evolves the function
\begin{equation}
  g = h - \frac{qF_0 \gyroavg{\varphi}{R}}{T_0}
  \label{def:g}
\end{equation}
according to \eqref{eq:gk} and \eqref{eq:qn}.
In order to make a 4D simulation using the 5D code, we used only 3 grid points along the field line (in $z$) with a very long box size and a periodic boundary condition.
Starting from the homogeneous initial condition along $z$, we confirmed that no structure developed in this direction.

The code uses a Fourier pseudo-spectral scheme for real space dimensions perpendicular to the mean magnetic field 
and Legendre collocation scheme in the two-dimensional velocity space integration.
The velocity-space grid is taken with respect to 
particle energy $\varepsilon = v^2$ and $\lambda = \vperp^2/\varepsilon$, which makes our grid points radially distributed in the $\vpar$-$\vperp$ plane \cite{Barnes-PoP10}.
For the collision operator we use a conservative finite difference scheme~\cite{Barnes-PoP09}.

Time integration is made using the 3rd order Adams-Bashforth scheme for the nonlinear term.
The linearized collision operator is treated by the first-order implicit Euler scheme 
with Sherman-Morrison formula for the moments-conserving corrections~\cite{Barnes-PoP09,Numerical Recipes}.

AstroGK is written in Fortran 95 and is parallelized using MPI.
Our biggest run (D2 in Table \ref{run table}) cost about 36 wall-clock hours using 8,192 processor cores.

\subsection{Initial condition}

Here we describe the setup of the direct cascade simulations.
We use the simple straight slab geometry in a homogeneous background in the box size $L_x = L_y = 2 \pi \rho$.
A series of decaying turbulence simulations were carried out, with the initial condition put in $|k_x \rho|, |k_y \rho| = 2$ scale:
\begin{align}
  g_{\rm init} = g_0 \left[ \cos \frac{2x}{\rho} + \cos \frac{2y}{\rho} 
    + \chi(x,y) \right] F_0.
  \label{eq:ginit}
\end{align}
where $F_0$ is a Maxwellian.
Here, $g_0$ is a constant and $\chi(x,y)$ is small-amplitude white noise superimposed on all Fourier harmonics.
From \eqref{eq:qn} and \eqref{def:g}, we can calculate $\varphi_{\rm init}$.
The initial condition \eqref{eq:ginit} is unstable to the kinetic version of the Kelvin-Helmholtz (KH) instability, which self-consistently evolves into turbulence.
It is a completely autonomous system and there is no entropy input during the time evolution.
We emphasize that there is no non-Maxwellian velocity-space structure in the initial condition.
All fine structure later observed in the velocity space is created from nonlinear interaction described in \sect{sec:equations}\null.

The results reported below were obtained in the series of runs indexed in Table~\ref{run table}, where $N_x\times N_y$ is number of collocation points in the position space and $N_\varepsilon\times 2 N_\lambda$ is the number of grid points in velocity space --- the factor of 2 corresponds to the sign of $\vpar = \pm\sqrt{\varepsilon(1-\lambda)}$. 
\begin{table}[bt]
  \caption{
    Index of the runs described in \sect{sec:direct}.
    Dimensionless number $D$ is given for the initial stage of
    developed turbulence ($t/\tau_{\rm init} = 10$), and
    cutoff wave number is estimated by \eqref{eq:cutoff}
    (see \sect{sec:evolution} and \fig{fig:do}).
  }
  \begin{center} \begin{tabular}{ccccccc}
      \hline \hline
      Run & $\nu \tau_{\rm init}$ & $D$ & $\kperpc \rho$ &
        $N_x \times N_y$ & $N_{\varepsilon} \times 2N_{\lambda}$ \\ \hline
      A & $9.3 \cdot 10^{-3}$  & $32$ & $16$ & $64^2$ & $32^2$ \\
      B & $5.6 \cdot 10^{-3}$  & $48$ & $20$ & $64^2$ & $32^2$ \\
      C1 & $1.9 \cdot 10^{-3}$ & $118$ & $35$ & $128^2$ & $16^2$ \\
      C2 & $1.9 \cdot 10^{-3}$ & $118$ & $35$ & $128^2$ & $64^2$ \\
      C3 & $1.9 \cdot 10^{-3}$ & $118$ & $35$ & $128^2$ & $128^2$ \\
      D1 & $7.4 \cdot 10^{-4}$ & $440$ & $77$ & $128^2$ & $128^2$ \\
      D2 & $7.4 \cdot 10^{-4}$ & $440$ & $77$ & $256^2$ & $128^2$ \\
      \hline \hline
  \end{tabular} \end{center}
  \label{run table}
\end{table}
For each collision frequency, $D$ is calculated using the turnover time taken from the simulation (see \sect{sec:evolution} and \fig{fig:do}).
The cutoff scale is listed for each case here, and we determine the grid resolution so that it resolves the smallest position- and velocity-space structures.

\subsection{Time evolution}
\label{sec:evolution}
\begin{figure}[bt]
  \centerline{
    \includegraphics[height=\figsize,angle=270]{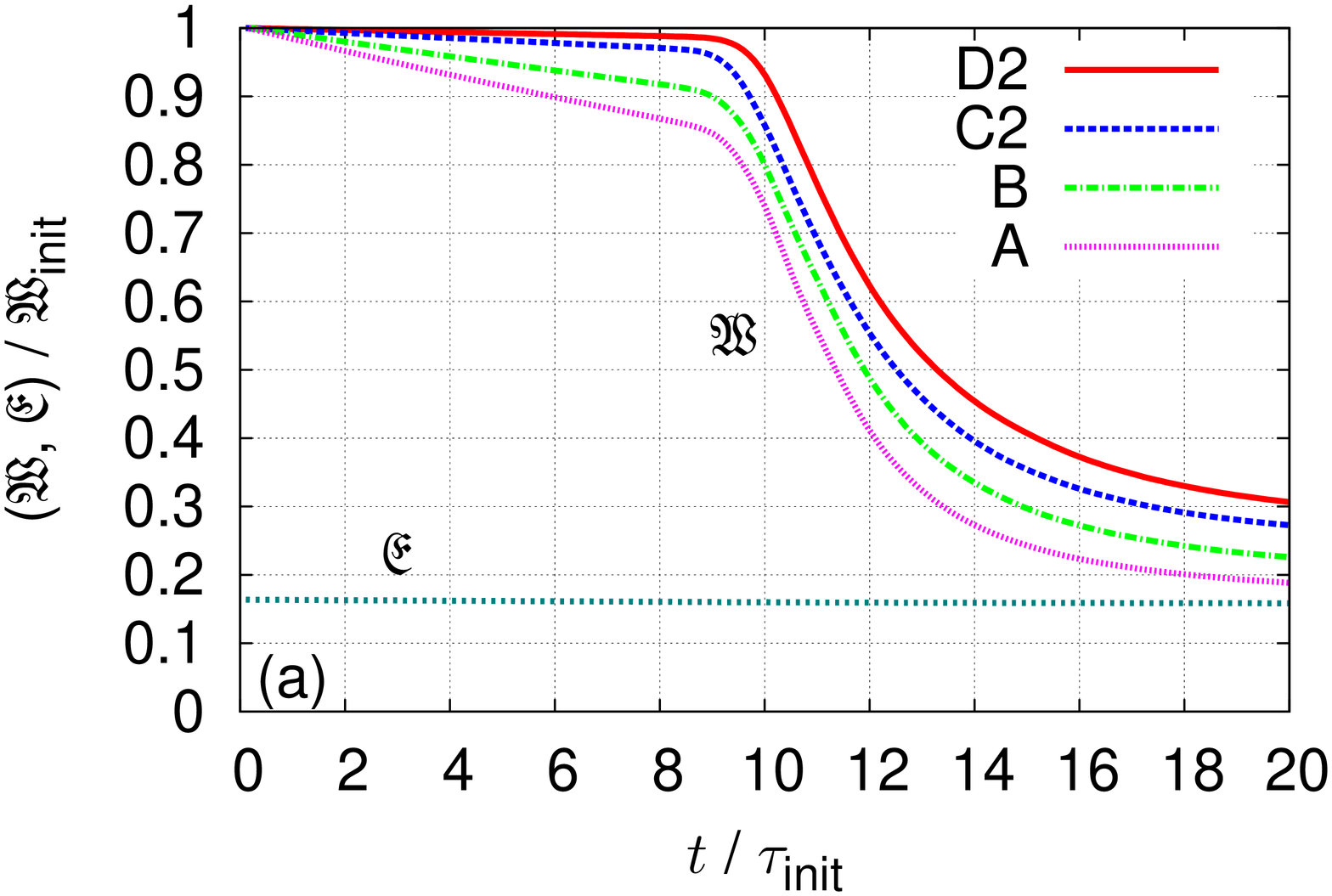}
  }
  \centerline{
    \includegraphics[height=\figsize,angle=270]{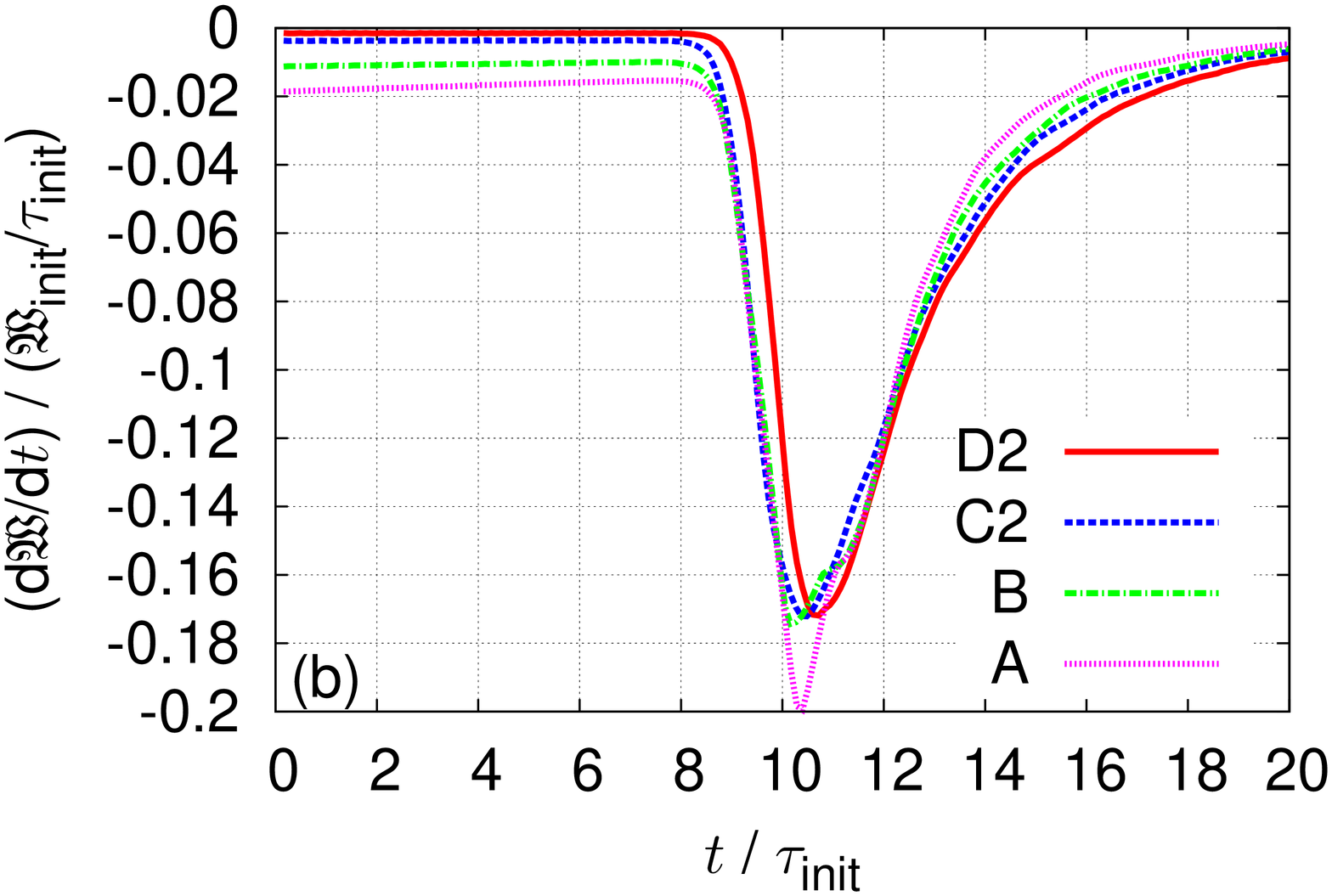}
  }
  \caption{
    Time evolution of (a) $\gt{W}$ and $\gt{E}$ [\eqref{eq:wes} and \eqref{eq:w4d}],
    and (b) rate of change of $\gt{W}$, normalized to initial $\gt{W}$.
    Labels A--D2 correspond to runs indexed in Table \ref{run table}.
    Evolution of $\gt{E}$ does not differ among runs significantly,
    and is represented by the run D2.
  }
  \label{fig:invariants}
\end{figure}
\begin{figure}[bt]
  \centerline{
    \includegraphics[height=\figsize,angle=270]{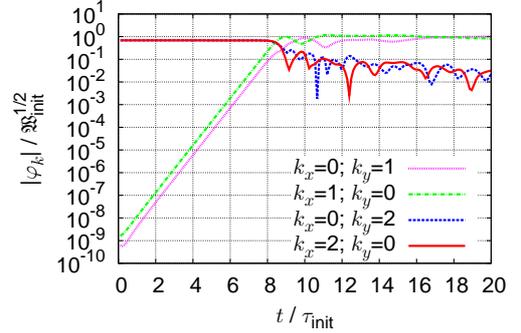}
  }
  \caption{
    Time evolution of several lower Fourier modes $\phik$ from the run D2
    (See Table \ref{run table}).
  }
  \label{fig:instability}
\end{figure}
Time evolution of the collisionless invariants and the amplitude of several lower Fourier modes $|\phik|$ are shown in \figs{fig:invariants}{fig:instability}, where $\tau_{\rm init}$ is the initial turnover time of the gyroaveraged \exb motion defined by
\begin{equation}
  \tau_{\rm init} = \frac{2 \pi B_0}{ck_xk_y ||\gyroavg{\varphi_{\rm init}}{R}||},
  \label{eq:tau}
\end{equation}
with $|k_x\rho|,|k_y\rho|=2$ and
\begin{equation}
  ||\gyroavg{\varphi}{R}|| = \left[ \frac{1}{n_0} \iint
    |\gyroavg{\varphi}{R}|^2 F_0 \, d\Vec{v} \, d\Vec{R} \right]^{1/2}.
  \label{typical potential}
\end{equation}
In \fig{fig:invariants}, the decrease of $\gt{W}$ corresponds to the creation of entropy since $\gt{W}$ is proportional to the negative of the perturbed entropy.
Initially, while instability grows ($t/\tau_{\rm init} \lesssim 8$, see also \fig{fig:instability}), $\gt{W}$ decays slowly with a rate proportional to the collision frequency $\nu$.
Then as the instability saturates around $t/\tau_{\rm init} \simeq{} 9$, $\gt{W}$ starts to decay very rapidly.
This rapid decay of $\gt{W}$ suggests nonlinear creation of small scale structures in velocity space as well as in position space.
Some time after that ($t/\tau_{\rm init} \gtrsim 20$), $\gt{W}$ starts to decay exponentially suggesting the end of turbulence.
The constancy of $\gt{E}$ implies the presence of an inverse cascade \cite{Plunk-JFM}, which will be discussed elsewhere.

Time evolution of the rate of change of $\gt{W}$ is shown in \fig{fig:invariants}(b).
Note that the decay of $\gt{W}$ is accounted for perfectly by collisional entropy production --- i.e.\ entropy balance \cite{WatanabeSugama-PoP04} is satisfied.
This is guaranteed under the controlled velocity-space numerical scheme \cite{Barnes-PoP10}.
In all runs except for the one with the smallest $D$ (run A in Table \ref{run table}; $D=32$), the rate of change of $\gt{W}$ is almost identical in the nonlinear stage ($t/\tau_{\rm init} \gtrsim 9$).
The slight horizontal shift of the lines is due to the different timing of linear mode saturation because of the random-noise admixture in the initial conditions.
This plot suggests that the asymptotic amount of dissipation is finite as collision frequency tends to zero, so the dissipation rate tends to a value asymptotically independent of the collisionality, but rather determined by the nonlinear dynamics of the turbulence.
Also, the falling phase of $d\gt{W}/dt$ corresponds to the development of the turbulence, and we may say that the turbulence has fully developed by $t/\tau_{\rm init} = 10$.
In the following sections, we will concentrate on the asymptotic cases, which have similar rates of change of $\gt{W}$ (runs B--D).

The dimensionless number $D$ is defined as follows.
As is seen in \fig{fig:instability}, the simple use of a single representative wave number $\kperp \rho = 2$ yields significant fluctuation of $D$ which is not necessarily physical.
On the other hand, the $\kperp \rho = 1$ component is created by the KH instability as a part of the inverse cascade, and is also irrelevant for characterizing direct cascade.
Thus we introduced a definition of the turnover time at scale $\ell = \pi \rho$, denoted by $\tau_{\pi\rho}$, which uses all $\kperp \rho \geq 2$ components of the fluctuating potential.
\begin{figure}[bt]
  \centerline{
    \includegraphics[height=\figsize,angle=270]{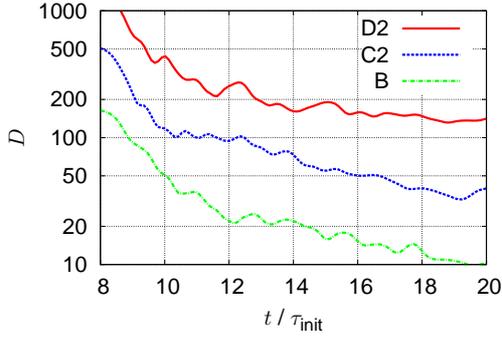}
  }
  \caption{
    Time evolution of the dimensionless number $D$ [see \eqref{dimensionless number}].
    Labels B--D2 correspond to runs indexed in Table \ref{run table}.
    Values of $D$ in Table \ref{run table} is evaluated at $t/\tau_{\rm init} = 10$.
  }
  \label{fig:do}
\end{figure}
Using $\tau_{\pi \rho}$, the time evolution of $D$ is shown in \fig{fig:do}.
The values are shown for three resolved cases from $t/\tau_{\rm init} \geq 8$ since the turbulent spectra start to develop after that.
At $t / \tau_{\rm init} = 10$ the dimensionless number $D$ takes values shown for each run in Table \ref{run table}\null.
The cutoff wave number is then obtained from
\begin{equation}
  \kperpc \rho = \alpha D^{3/5}.
  \label{eq:cutoff}
\end{equation}
The value $\alpha = 2$ corresponds to the particular initial condition and domain size used here.

\subsection{Wave-number spectra}
\label{sec:k-spectra}

\begin{figure}[bt]
  \centerline{
    \includegraphics[height=\figsize,angle=270]
      {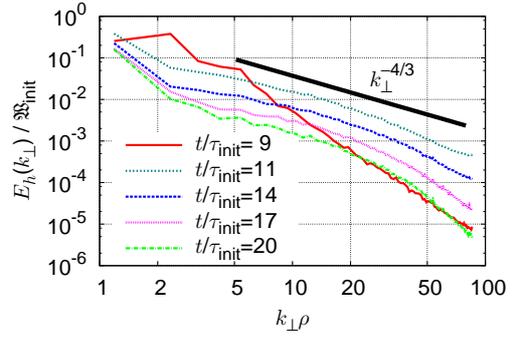}
  }
  \caption{
    Time evolution of the wave-number spectra of $E_h(\kperp)$ from the run D2
    (see Table~\ref{run table}).
  }
  \label{fig:k-spectra-evol}
\end{figure}
We first show the time evolution of the wave-number spectra of $E_h(\kperp)$ averaged over wave-number shell [defined by \eqref{def:h-spectra}] in \fig{fig:k-spectra-evol}.
At $t/\tau_{\rm init}=9$ the spectrum shows an overall steeper slope as it has not yet entered the fully developed turbulence stage.
From $t/\tau_{\rm init} = 11$ to $14$ the spectra show a self-similar shape consistent with the inertial-range scaling \eqref{eq:spectra}.
The spectra at $t/\tau_{\rm init}=17$ and $20$ clearly have an inertial range;
the steeper falloff at large $\kperp$ suggests a dissipation range.
Note that the implied cutoff scale is consistent with the estimate from the time-evolving dimensionless number $D$ [see \fig{fig:do} and \eqref{eq:cutoff}].

\begin{figure}[bt]
  \centerline{
    \includegraphics[height=\figsize,angle=270]
      {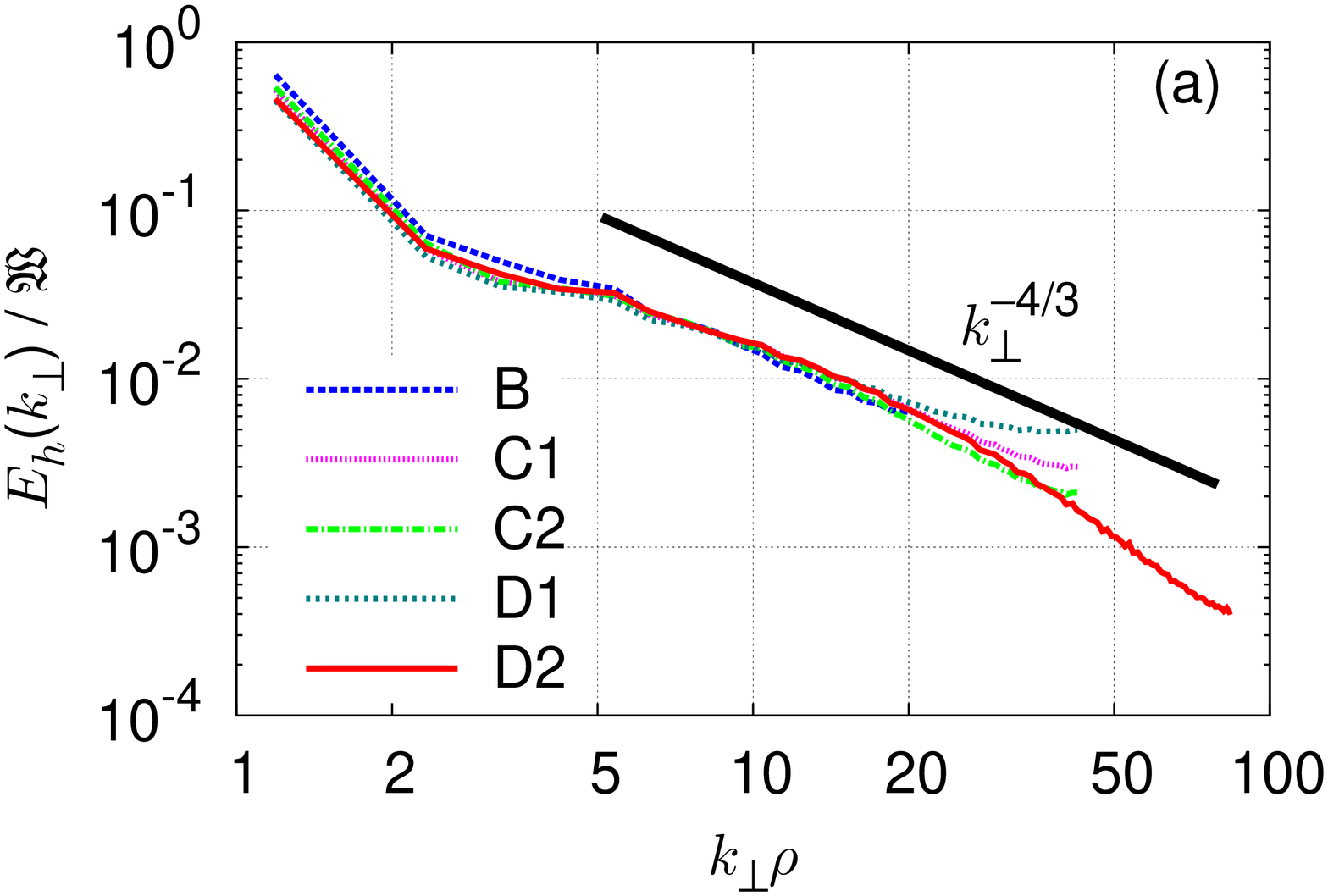}
  }
\centerline{
    \includegraphics[height=\figsize,angle=270]
      {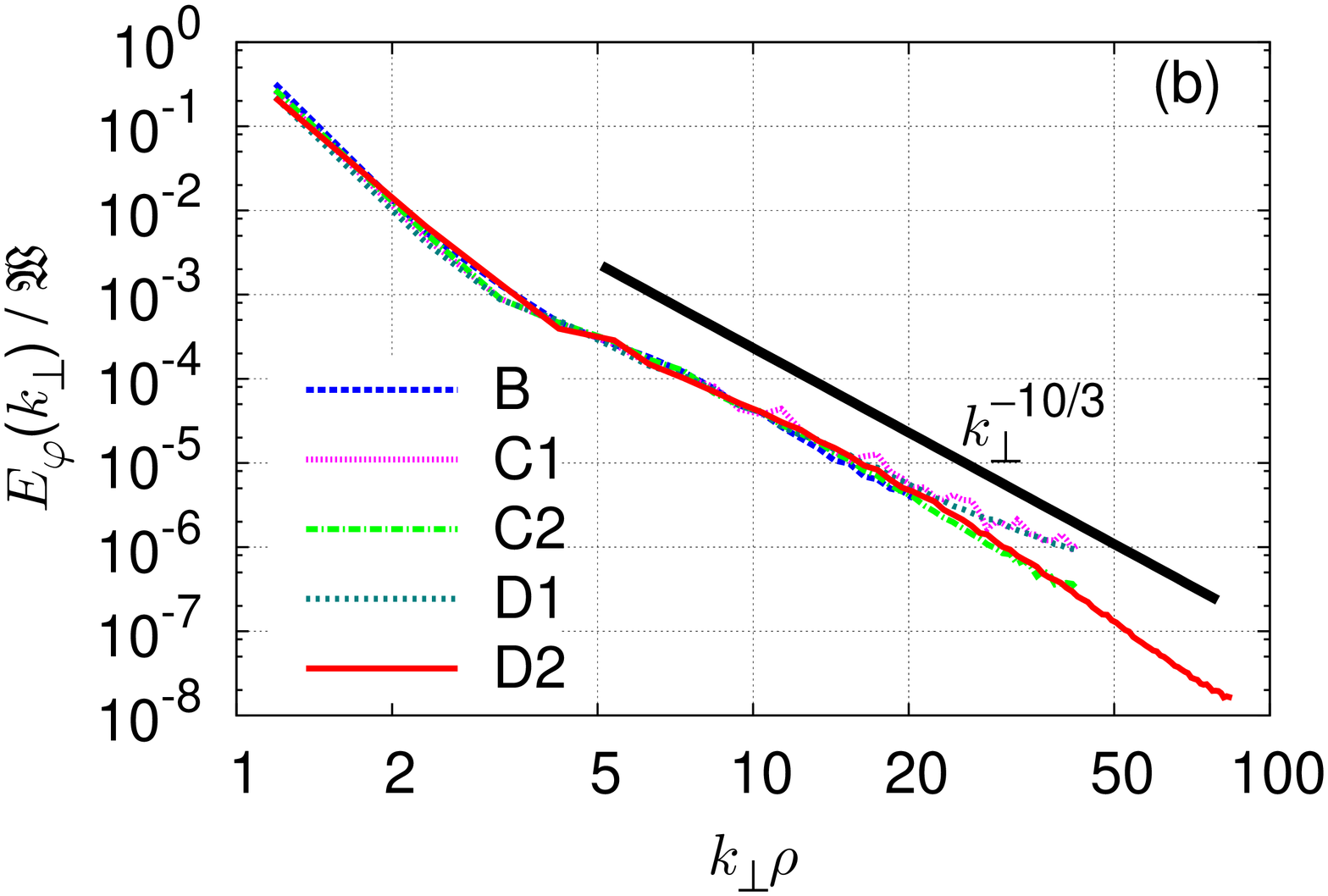}
  }
  \caption{
    Time-averaged normalized wave-number (Fourier) spectra (a) $E_h(\kperp)/\gt{W}$
    and (b) $E_{\varphi}(\kperp)/\gt{W}$ [cf.\ \eqref{eq:spectra}] for the runs
    indexed in Table~\ref{run table}.
    Theoretically predicted slopes are given for comparison. 
  }
  \label{fig:k-spectra}
\end{figure}
Figure \ref{fig:k-spectra} shows the wave-number spectra of $E_h(\kperp)$ and $E_{\varphi}(\kperp)$ [see \eqref{def:h-spectra} and \eqref{def:phi-spectra}] of the asymptotic runs (B--D in Table \ref{run table}) which are averaged over wave-number shell and normalized by the value of total $\gt{W}$ at each time, and then averaged over time $10 \leq t / \tau_{\rm init} \leq 15$.
It is hard to identify the dissipation range as the time average is taken
over the period when the cutoff is near the highest $\kperp$ end so that
inertial range can be taken as wide as possible.
Unclear dissipation range in the time-averaged spectra is acceptable
as exponential fall-off in this regime implies exponentially small
dissipation apart from the cutoff.
Three resolved cases B, C2 and D2 lie on top of one another and are consistent with the theoretical slopes ($E_h \propto \kperp^{-4/3}$ and $E_{\varphi} \propto \kperp^{-10/3}$).
In contrast, the under-resolved runs in the velocity space (C1) and in the position space (D1) have shallower slopes than others.
With the increase of $D$ the dissipation cutoff extends to smaller scales, thus higher resolution is required both in position and velocity space.
The simulation shows a consistent result with the scaling theory including the dissipation cutoff.

\subsection{Velocity-space spectra}
In order to characterize the velocity-space structure quantitatively, Plunk \textit{et al}.\ \cite{Plunk-JFM} proposed a dual space variable $p$ to the velocity variable $\vperp$ and introduced a 2D spectrum in $(\kperp,p)$ space:
\begin{equation}
  \hat{E}_g(\kperp,p) = \sum_{|\Vec{k}_\perp| = \kperp} p |\hat g_{\Vec{k}}(p)|^2,
  \label{eq:Hankel}
\end{equation}
where $\hat g_{\Vec k}(p) = \int J_0(p \vperp) g_{\Vec{k}}(\Vec{v}) \, d\Vec{v}$ is the Hankel transform.

\begin{figure}[bt]
  \centerline{
    \includegraphics[width=6cm]
      {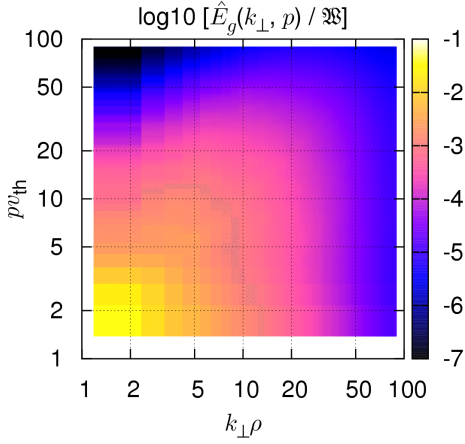}
  }
  \caption{
    Time-averaged normalized 2D spectrum $\hat{E}_g(\kperp,p)$
      [defined by \eqref{eq:Hankel}] taken from the run D2
    (See Table \ref{run table}).
  }
  \label{fig:2d-spectrum}
\end{figure}
Figure \ref{fig:2d-spectrum} shows the time average of the normalized 2D spectrum $\hat{E}_g(\kperp,p)$ taken from the run D2 (see Table \ref{run table}).
We first note that there is a gap in the low-$\kperp$, high-$p$ region.
This gap appears because high-$p$ component is not easily created with a low-$\kperp$ mode, while the high-$\kperp$ component can be created even for low-$p$ as Navier-Stokes turbulence does with Maxwellian velocity profile.
This is likely due to the fact that nonlinear phase mixing acts more strongly at sub-Larmor scales.
The spectral contour develops along the diagonal $p \vth = \kperp \rho$,
which reflects the fact that the correlation of position- and velocity-space structure follows our conjecture \eqref{correlation} very well.
Thus, comparable resolution in velocity and position space is required, especially when we investigate smaller scales than the Larmor radius.

\begin{figure}[bt]
  \centerline{
    \includegraphics[height=\figsize,angle=270]{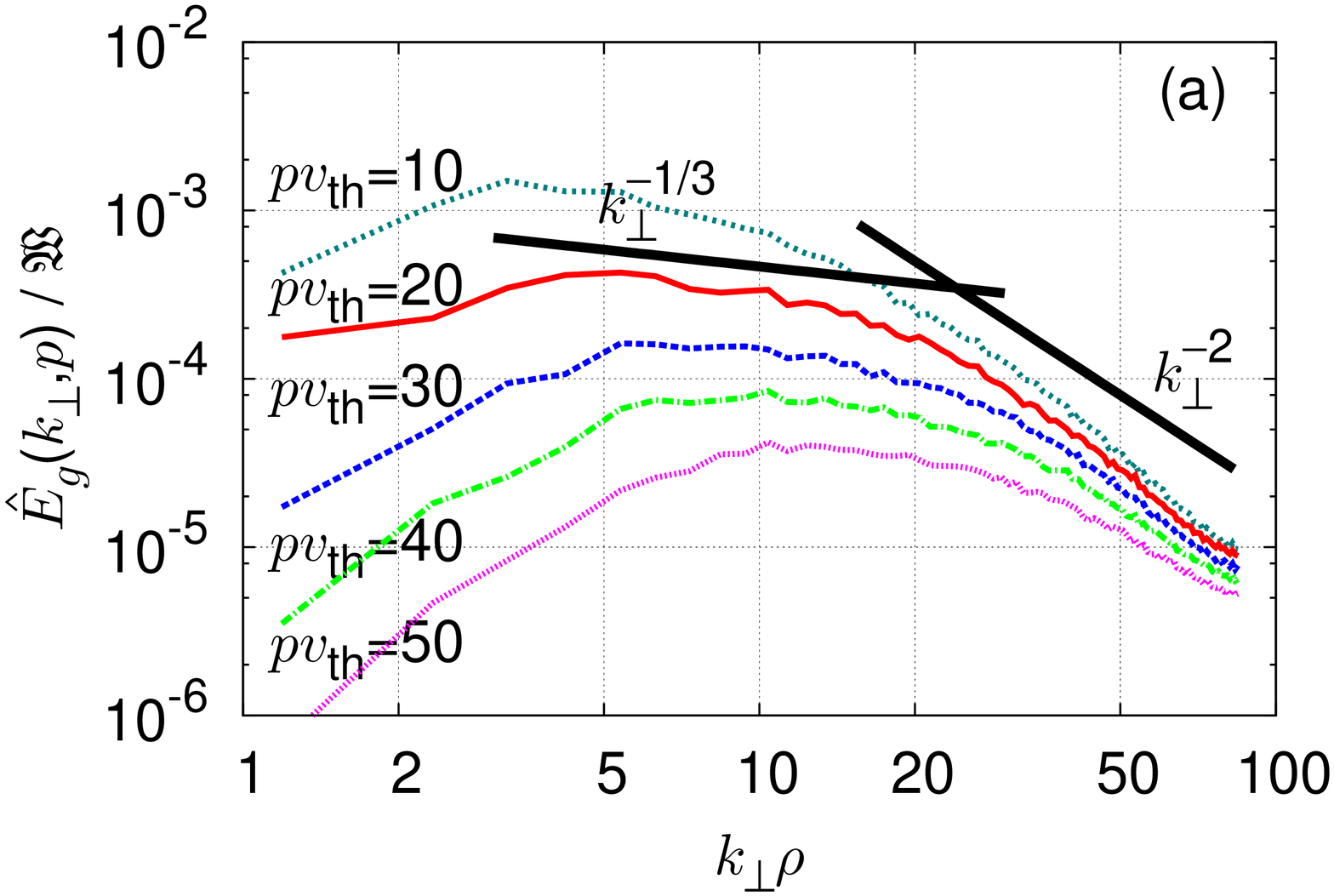}
  }
  \centerline{
    \includegraphics[height=\figsize,angle=270]{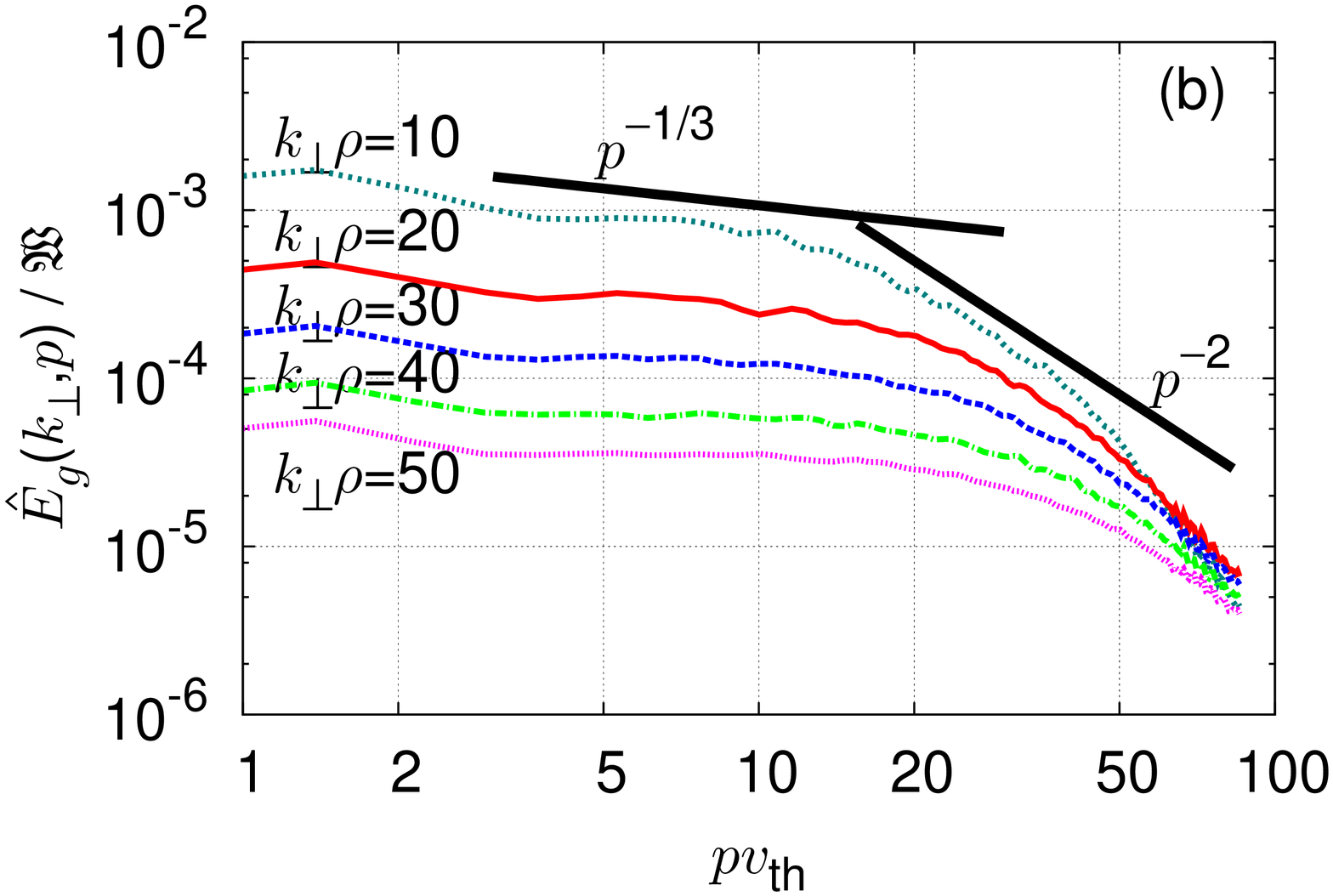}
  }
  \caption{
    Averaged asymptotic spectra from the run D2 
    (See Table \ref{run table})
    (a) as a function of $\kperp$ for various fixed $p$
    and (b) as a function of $p$ for various fixed $\kperp$.
  }
  \label{fig:asymptotic spectra}
\end{figure}
In Ref.\ \cite{Plunk-JFM} the asymptotic spectra are derived for
$\kperp \rho \gg p \vth$ and $\kperp \rho \ll p \vth$ limits, namely
\begin{equation}
  \hat{E}_g(\kperp,p) \propto \begin{cases}
    \kperp^{-2} p^{-1/3} & (\kperp \rho \gg p \vth) \\
    p^{-2} \kperp^{-1/3} & (\kperp \rho \ll p \vth)
    \label{eq:asymptotic spectra}
  \end{cases}.
\end{equation}
Figure \ref{fig:asymptotic spectra} shows various (a) horizontal and (b) vertical slices of \fig{fig:2d-spectrum}.
Depending on the value of $p$ in \fig{fig:asymptotic spectra}(a), we observe a slope approaching $\kperp^{-2}$ in the high-$\kperp$ regime.
On the other hand, the $\kperp^{-1/3}$ slope for $\kperp \rho \ll p \vth$ seems difficult to obtain, reflecting the gap in the low-$\kperp$, high-$p$ region seen in \fig{fig:2d-spectrum}.
Figure \ref{fig:asymptotic spectra}(b) shows a rather good slope consistent with that predicted for the low-$p$ limit (${} \propto p^{-1/3}$) even for the smallest $\kperp$ in the figure.
The high-$p$ slope is steeper than the prediction \eqref{eq:asymptotic spectra} for the $\kperp \rho = 10$ case because of the remnant of the gap in \fig{fig:2d-spectrum}.
The agreement with the asymptotic expectation becomes good for $\kperp \rho \gtrsim 30$ even though we don't have a wide asymptotic range in this region.
The overall spectra show a reasonable symmetry in $\kperp$ and $p$ except for the structure coming from the gap, which indicates consistency with the theoretical prediction \eqref{eq:asymptotic spectra}.

\begin{figure}[bt]
  \centerline{
    \includegraphics[height=\figsize,angle=270]{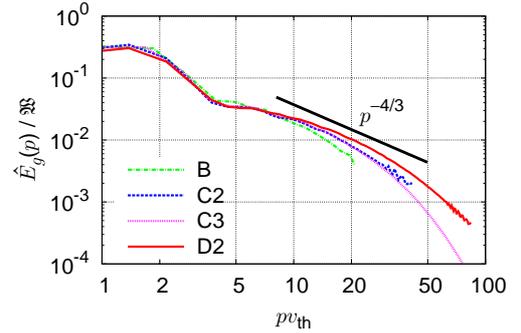}
  }
  \caption{
    Time-averaged normalized velocity-space (Hankel) spectrum $\hat{E}_g(p)
    = \int \hat{E}_g(\kperp,p) \, d\kperp$ [see also \eqref{eq:Hankel}]
    for the runs in Table \ref{run table}.
  }
  \label{fig:v-spectra}
\end{figure}
Figure \ref{fig:v-spectra} shows the time-averaged spectra in the Hankel space $\hat{E}_g(p) = \int \hat{E}_g(\kperp,p) \, d\kperp$ normalized to $\gt{W}$ at each time.
The theoretical expectation is \cite{Plunk-JFM}
\begin{equation}
  \hat{E}_g(p) \propto p^{-4/3},
  \label{eq:v-spectra}
\end{equation}
which logically follows from the first spectrum in \eqref{eq:spectra} and $p \vth \sim \kperp \rho$ [equivalent to \eqref{correlation}].
The numerical result again shows approximate consistency with the theoretical prediction and confirms that small-scale structure is formed in the velocity space.
The large hump in the low-$p$ regime ($p \lesssim 3$) is due to long-wave-length modes, which have significantly larger amplitudes than the rest.
These long-wave-length modes have velocity-space structure close to Maxwellian, whose Hankel transform yields $\int J_0(p\vperp) \, e^{-\vperp^2/\vth^2} \vperp \, d\vperp / \vth^2 = e^{-p^2\vth^2/4}/2$.
The gradual steepening at the high-$p$ region may be related to the gap in \fig{fig:2d-spectrum} (Notice that \fig{fig:v-spectra} corresponds to taking the horizontal sum of \fig{fig:2d-spectrum}).
The wiggles in the high-$p$ end of the runs C2 and D2 come from the slight lack of resolution, which goes away by increasing the velocity resolution as the high velocity-resolution run C3 shows.
This increase of the velocity-space resolution does not affect the wave-number spectra, while failure to resolve the scaling regime does (see line C1 in \fig{fig:k-spectra}).

\subsection{Entropy transfer}
In the scaling theory (\sect{sec:theory}) we assumed local-scale interaction by following Kolmogorov's argument.
The applicability of such an assumption may be directly investigated in the numerical simulation.

In order to make this diagnostic we have introduced a Fourier filtered function \cite{AlexakisMininniPouquet-PRE05} defined by
\begin{equation}
  g_K(\Vec{R},\vperp) := \sum_{\Vec{k} \in {\cal K}} g_{\Vec{k}}(\vperp)
    e^{{\rm i} \Vec{k} \cdot \Vec{R}},
\end{equation}
where ${\cal K} = \{ \Vec{k}: K\rho-1/2 \leq |\Vec{k}|\rho < K\rho+1/2 \}$.
The filtering is orthogonal, and $g_K$ denotes the component of the distribution function with scale $K^{-1}$.
Then the entropy amount contained in this scale obeys the evolution equation
\begin{equation}
  \frac{d}{dt} \int \frac{|g_K|^2}{2 F_0} \, d\Vec{R}
    \, d\Vec{v} = \sum_Q T_F(K,Q) - \mathrm{collisions},
\end{equation}
where $T_F(K,Q)$ denotes the nonlinear transfer of entropy from scale $Q^{-1}$ to $K^{-1}$
\begin{equation}
  T_F(K,Q) := - \iint \vperp \frac{g_K \vexb \cdot \nabla g_Q}{F_0} \,
    d\Vec{R} \, d\vperp,
    \label{eq:ktrans}
\end{equation}
which is by definition \change{antisymmetric} with respect to the exchange of two arguments.

\begin{figure}[bt]
  \centerline{
    \includegraphics[width=\figsize]{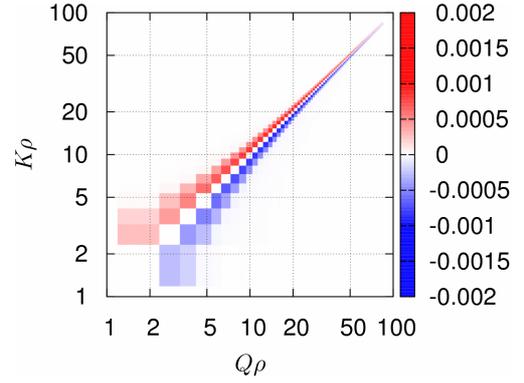}
  }
  \vskip -2mm
  \caption{
    Time-averaged normalized entropy transfer function $T_F(K,Q)$
    [see also \eqref{eq:ktrans}] for the run D2 in Table \ref{run table}.
  }
  \label{fig:ktrans}
\end{figure}

The numerical result obtained from the run D2 (see Table \ref{run table}) is shown in \fig{fig:ktrans}, which is normalized by $\gt{W}$ at each time and averaged over $10 \leq t / \tau_{\rm init} \leq 15$.
Figure \ref{fig:ktrans} shows a remarkable locality of the interaction, which is clearly seen even at each time.
The color denotes the direction of the entropy transfer, which indicates that the entropy is transferred from large scales to small scales.

The scaling theory does not require locality in velocity space scales, however the velocity-space transfer function may still give some useful information.
It may be monitored by Hankel filtered function defined by
\begin{equation}
  g_P(\vperp) := \int_{p \in {\cal P}} \hat{g}_{\Vec{k}}(p) J_0(p\vperp) p \, dp,
  \label{eq:Hankel filter}
\end{equation}
where ${\cal P} = \{ p: P\vth -1/2 \leq p\vth < P\vth+1/2 \}$.
Then, similarly to the Fourier filtering,
entropy transfer function in Hankel space is described by
\begin{equation}
  T_H(P,S) := - \iint \check{g}_P \vexb \cdot \nabla \check{g}_S \vperp
  \, d\vperp \, d\Vec{R},
  \label{eq:ptrans}
\end{equation}
where $\check{g}(\vperp) := g(\vperp) / \sqrt{F_0}$ is introduced to make entropy a bilinear form, and $T_H(P,S)=-T_H(S,P)$.

\begin{figure}[bt]
  \centerline{
    \includegraphics[width=\figsize]{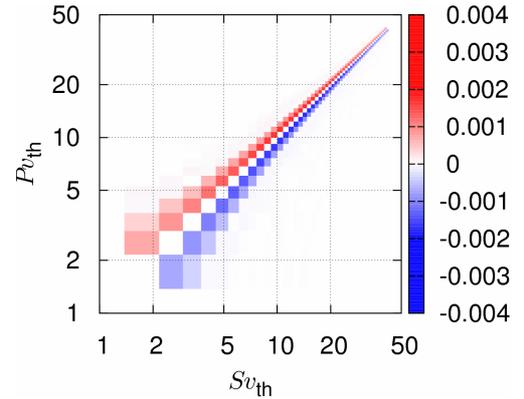}
  }
  \vskip -2mm
  \caption{
    Time-averaged normalized entropy transfer function $T_H(P,S)$
    [see also \eqref{eq:ptrans}] for the run C2 in Table \ref{run table}.
  }
  \label{fig:ptrans}
\end{figure}

Figure \ref{fig:ptrans} shows the normalized, time-averaged transfer function in the velocity space.
For numerical purposes the integral in \eqref{eq:Hankel filter} is approximated by a single representative Hankel mode.
Nevertheless the transfer function is localized along the diagonal very well, which, together with \fig{fig:ktrans}, may suggest the entropy transfer along the diagonal in $(\kperp,p)$-plane (see \fig{fig:2d-spectrum}).

\section{Summary}
\label{sec:summary}

We presented electrostatic, decaying turbulence simulations for weakly collisional, magnetized plasmas using the gyrokinetic model in 4D phase space (two position-space and two velocity-space dimensions; the extension to three spatial dimensions is left for future work).
Landau damping was removed from the system by ignoring variation along the background magnetic field.

Nonlinear interactions introduce an amplitude-dependent perpendicular phase mixing of the gyrophase-independent part of the perturbed distribution function and create structure in $v_\perp$ which is finer for higher $\kperp$ (see \fig{fig:2d-spectrum}).
We found that the wave-number (Fourier) spectra of the perturbed distribution function and the resulting electrostatic fluctuations at sub-Larmor scales agreed well with theoretical predictions based on the interpretation of the nonlinear phase mixing as a cascade of entropy in phase space (see Figs.\ \ref{fig:k-spectra} and \ref{fig:asymptotic spectra}) \cite{Tatsuno-PRL09,Alex,Plunk-JFM}. 
The velocity-space (Hankel) spectra show a rough consistency with the theoretical scaling, although the agreement is not as good as that of the wave-number spectra (see Figs.\ \ref{fig:v-spectra} and \ref{fig:2d-spectrum}).

We introduced a dimensionless number (analogous to Reynolds number) that characterized the scale separation between the thermal Larmor scale and the collisional cutoff in phase space [see \eqref{dimensionless number}], and showed that this number correctly predicted the resolution requirements for nonlinear gyrokinetic simulations (see Table \ref{run table}).
We also showed the trend that the entropy generation rate (or irreversibility) is independent of the collision frequency in the asymptotic limit of the weak collisionality (see \fig{fig:invariants}).
Finally we presented diagnostics of nonlinear transfer functions and showed that local-scale cascade of entropy is supported very well in both Fourier and Hankel spaces.

We note that there are, in general, entropy cascades for each plasma species.
Equations for the gyrokinetic turbulence at and below the electron Larmor scale are mathematically similar to the model simulated here and identical arguments apply \cite{Alex,Plunk-JFM}.
Similar considerations are also possible for ion-scale electromagnetic turbulence \cite{Alex} and for minority ion species (with some differences to be discussed elsewhere).

The structure of the small scales in phase space that we have presented is likely to be a universal feature of magnetized plasma turbulence.
Understanding it theoretically and diagnosing it numerically is akin to the inertial-range studies for Kolmogorov turbulence, extended to the kinetic phase space.
One should expect rich and interesting physics to emerge and it is likely that, just like in the case of fluid turbulence, predicting large-scale dynamics will require effective models for the small-scale cascade.
An immediate physical implication of the existence of the entropy cascade is a turbulent heating rate independent of collisionality in weakly collisional plasmas.

\section*{Acknowledgments}
TT, WD, RN and GGP thank the Leverhulme Trust Network for Magnetised Plasma Turbulence for travel support.
AAS was supported by STFC.
Numerical computations were performed at NERSC, NCSA, and TACC.


\begin{thebibliography}{99}

\bibitem{Kolmogorov}
  A.~N.~Kolmogorov, Dokl.\ Akad.\ Nauk SSSR \textbf{30}, 299 (1941);
  \textit{ibid}.\ \textbf{32}, 16 (1941)
  [Proc.\ Roy.\ Soc.\ London A \textbf{434}, 9 (1991);
  \textit{ibid}.\ \textbf{434}, 15 (1991)].

\bibitem{WatanabeSugama-PoP04}
  T.-H.~Watanabe and H.~Sugama, Phys.\ Plasmas \textbf{11}, 1476 (2004).

\bibitem{Idomura-PoP06}
  Y.~Idomura, Phys.\ Plasmas \textbf{13}, 080701 (2006).

\bibitem{Candy-PPCF07}
  J.~Candy, R.~E.~Waltz, M.~R.~Fahey \textit{et al}.,
  Plasma Phys.\ Control.\ Fusion \textbf{49}, 1209 (2007).

\bibitem{Bale-PRL05}
  S.~D.~Bale, P.~J.~Kellogg, F.~S.~Mozer \textit{et al}.,
  \prl \textbf{94}, 215002 (2005).

\bibitem{Howes-PRL08}
  G.~G.~Howes, W.~Dorland, S.~C.~Cowley \textit{et al}.,
  \prl \textbf{100}, 065004 (2008).

\bibitem{GarySaitoLi}
  S.~P.~Gary, S.~Saito, and H.~Li, 
    Geophys.\ Res.\ Lett., \textbf{35}, L02104 (2008);
  S.~Saito, S.~P.~Gary, H.~Li \textit{et al}.,
    Phys.\ Plasmas \textbf{15}, 102305 (2008).

\bibitem{Sahraoui-PRL09}
  F.~Sahraoui, M.~L.~Goldstein, P.~Robert \textit{et al}.,
  \prl \textbf{102}, 231102 (2009).

\bibitem{Alexandrova-PRL09}
  O.~Alexandrova, J.~Saur, C.~Lacombe \textit{et al}.,
  \prl \textbf{103}, 165003 (2009).

\bibitem{Boltzmann}
  L.~Boltzmann, Sitsungsber.\ Akad.\ Wiss.\ \textbf{66}, 275 (1872)
  [in \textit{Kinetic Theory 2} (Pergamon, Oxford, 1966) 88].

\bibitem{Helander}
  P.~Helander and D.~J.~Sigmar,
  \textit{Collisional Transport in Magnetized Plasmas}
  (Cambridge Univ., Cambridge, 2002).

\bibitem{Tatsuno-PRL09}
  T.~Tatsuno, W.~Dorland, A.~A.~Schekochihin \textit{et al}.,
  \prl \textbf{103}, 015003 (2009).

\bibitem{Fjortoft-Tellus53}
  R.~Fj{\o}rtoft, Tellus \textbf{2}, 225 (1953).

\bibitem{Kraichnan-PoF67}
  R.~H.~Kraichnan, Phys.\ Fluids \textbf{10}, 1417 (1967).

\bibitem{HasegawaMima-PoF78}
  A.~Hasegawa and K.~Mima, Phys.\ Fluids \textbf{21}, 87 (1978).

\bibitem{Horton-PoP00}
  W.~Horton, P.~Zhu, G.~T.~Hoang \textit{et al}.,
  Phys. Plasmas \textbf{7}, 1494 (2000).

\bibitem{Catto-PP78}
  P.~Catto, Plasma Phys.\ \textbf{20}, 719 (1978).

\bibitem{AntonsenLane-PoF80}
  T.~M.~Antonsen and B.~Lane, Phys.\ Fluids \textbf{23}, 1205 (1980).

\bibitem{FriemanChen-PoF82}
  E.~A.~Frieman and L.~Chen, Phys.\ Fluids \textbf{25}, 502 (1982).

\bibitem{Howes-ApJ06}
  G.~G.~Howes, S.~C.~Cowley, W.~Dorland \textit{et al}.,
  Astrophys.\ J. \textbf{651}, 590 (2006).

\bibitem{Abel-PoP08}
  I.~G.~Abel, M.~Barnes, S.~C.~Cowley \textit{et al}.,
  Phys.\ Plasmas, \textbf{15}, 122509 (2008).

\bibitem{DorlandHammett-PoFB93}
  W.~Dorland and G.~W.~Hammett, Phys.\ Fluids B \textbf{5}, 812 (1993).

\bibitem{Krommes}
  J.~A.~Krommes, Phys.\ Plasmas \textbf{6}, 1477 (1999).

\bibitem{Hallatschek}
  K.~Hallatschek, \prl \textbf{93}, 125001 (2004);
  B.~D.~Scott, \arxiv{0710.4899} (2007).

\bibitem{Alex}
  A.~A.~Schekochihin, S.~C.~Cowley, W.~Dorland \textit{et al}.,
  Astrophys.\ J.\ Suppl.\ Ser.\ \textbf{182}, 310 (2009);
  A.~A.~Schekochihin, S.~C.~Cowley, W.~Dorland \textit{et al}.,
  Plasma Phys.\ Control.\ Fusion \textbf{50}, 124024 (2008).

\bibitem{Gustafson-PoP08}
   K.~Gustafson, D.~del-Castillo-Negrete, and W.~Dorland,
   Phys.\ Plasmas, \textbf{15}, 102309 (2008).

\bibitem{Plunk-JFM}
  G.~G.~Plunk, S.~C.~Cowley, A.~A.~Schekochihin \textit{et al}.,
  J. Fluid Mech., \change{in press (2010)}; \arxiv{0904.0243}.

\bibitem{CarterMaggs-PoP09}
  T.~A.~Carter and J.~E.~Maggs, Phys.\ Plasmas \textbf{16}, 012304 (2009).

\bibitem{Jakubowski-PRL02}
  M.~Jakubowski, R.~J.~Fonck, and G.~R.~McKee, \prl \textbf{89}, 265003 (2002).

\bibitem{Durst-PRL93}
  R.~D.~Durst, R.~J.~Fonck, J.~S.~Kim \textit{et al}.,
  \prl \textbf{71}, 3135 (1993).

\bibitem{Mazzucato-PRL08}
  E.~Mazzucato, D.~R.~Smith, R.~E.~Bell \textit{et al}.,
  \prl \textbf{101}, 075001 (2008).

\bibitem{JenkoDorland-PRL02}
  F.~Jenko and W.~Dorland, \prl \textbf{89}, 225001 (2002).

\bibitem{Barnes-PoP10}
  M.~Barnes, W.~Dorland, and T.~Tatsuno,
  Phys.\ Plasmas \change{\textbf{17}, 032106} (2010).

\bibitem{Barnes-PoP09}
  M.~Barnes, I.~G.~Abel, W.~Dorland \textit{et al}.,
  Phys.\ Plasmas \textbf{16}, 072107 (2009).

\bibitem{NumericalRecipes}
  W.~H.~Press, S.~A.~Teukolsky, W.~T.~Vetterling \textit{et al}.,
  \textit{Numerical Recipes} (Cambridge Univ., Cambridge, 1988).

\bibitem{AlexakisMininniPouquet-PRE05}
  A.~Alexakis, P.~D.~Mininni, and A.~Pouquet,
  Phys.\ Rev.\ E \textbf{72}, 046301 (2005).

\end{thebibliography}
\end{document}